\documentclass{emulateapj}

\usepackage{amsfonts}
\usepackage{float}
\usepackage{amsmath}
\usepackage{epsfig,floatflt}
\usepackage{subfigure}

\newcommand{\Bs}{\mathbf{s}}
\newcommand{\BS}{\mathbf{S}}
\newcommand{\Bn}{\mathbf{n}}
\newcommand{\BN}{\mathbf{N}}
\newcommand{\Bm}{\mathbf{m}}
\newcommand{\Bf}{\mathbf{f}}

\newcommand{\Bd}{\mathbf{d}}
\newcommand{\BC}{\mathbf{C}}
\newcommand{\BB}{\mathbf{B}}
\newcommand{\BW}{\mathbf{W}}

\newcommand{\id}{\mathbf{1}}
\newcommand{\Bx}{\mathbf{x}}
\newcommand{\By}{\mathbf{y}}

\begin{document}

\title{Estimation of Polarized Power Spectra by Gibbs sampling}
\date{July 5, 2006}

\author{D. L. Larson\altaffilmark{2}, H.\ K.\
  Eriksen\altaffilmark{3,4,5,6},  B.\ D.\
  Wandelt\altaffilmark{2,7}, K. M.
  G\'{o}rski\altaffilmark{5,6,8}, \\Greg Huey\altaffilmark{2}, J. B.
  Jewell\altaffilmark{5}, I.\ J.\ O'Dwyer\altaffilmark{5,6}}

\altaffiltext{1}{email: dlarson1@uiuc.edu}

\altaffiltext{2}{Department of Physics, University of Illinois,
  Urbana, IL 61801}

\altaffiltext{3}{Institute of Theoretical Astrophysics, University of
  Oslo, P.O.\ Box 1029 Blindern, N-0315 Oslo, Norway}

\altaffiltext{4}{Centre of
  Mathematics for Applications, University of Oslo, P.O.\ Box 1053
  Blindern, N-0316 Oslo}

\altaffiltext{5}{Jet Propulsion Laboratory, 4800 Oak
  Grove Drive, Pasadena CA 91109} 

\altaffiltext{6}{California Institute of Technology, Pasadena, CA
  91125} 

\altaffiltext{7}{Astronomy Department, University of Illinois at
  Urbana-Champaign, IL 61801-3080}

\altaffiltext{8}{Warsaw University Observatory, Aleje Ujazdowskie 4, 00-478 Warszawa,
  Poland}

\date{Received - / Accepted -}


\begin{abstract}
  Earlier papers introduced a method of accurately estimating the
  angular cosmic microwave background (CMB) temperature power spectrum
  based on Gibbs sampling.  Here we extend this framework to polarized
  data. All advantages of the Gibbs sampler still apply, and exact
  analysis of mega-pixel polarized data sets is thus feasible.  These
  advantages may be even more important for polarization measurements
  than for temperature measurements.  While approximate methods can
  alias power from the larger E-mode spectrum into the weaker B-mode
  spectrum, the Gibbs sampler (or equivalently, exact likelihood
  evaluations) allows for a statistically optimal separation of these
  modes in terms of power spectra.  To demonstrate the method, we
  analyze two simulated data sets: 1)~a hypothetical future CMBPol
  mission, with the focus on B-mode estimation; and 2)~a Planck-like
  mission, to highlight the computational feasibility of the method.
\end{abstract}


\section{Introduction}

Since the first detection of cosmic microwave background (CMB)
polarization (DASI; Kovac et al.\ 2002, Leitch et al.\ 2002) and
subsequent measurement of the temperature-gradient (TE) cross-power
spectrum by the Wilkinson Microwave Anisotropy Probe (WMAP; Kogut et
al.\ 2003), emphasis has shifted to the measurement and analysis of
the full polarization angular power spectra.  Many experiments
\citep{leitch:2005, readhead:2004, barkats:2005, montroy:2005,
page:2006} have improved on those early findings, producing
measurements with a considerable gain in raw sensitivity.  However, an
important concern with all such measurements is systematic errors,
including not only instrumental effects, observing strategy effects,
and astrophysical contaminants, but also statistical issues.  It is
essential to develop powerful and flexible data analysis tools to
extract the desired information from the raw data reliably.  In this
paper we progress towards this goal by extending the previously
introduced Gibbs sampling framework
\citep{jewell:2004, wandelt:2004, eriksen:2004} to polarization.

The scientific importance of CMB polarization power spectra is high.
For example, our current understanding of the optical depth,
amplitude, and scalar spectral index hinges on what we know about the
magnitude of the the low-$\ell$ temperature and polarization spectra
from the WMAP 3-year data \citep{page:2006}.  Also, a detection of
large scale B~modes would give a very exciting insight into primordial
gravitational waves.

Earlier Gibbs analyses of unpolarized CMB data were described by
\cite{wandelt:2004, odwyer:2004, eriksen:2004, eriksen:2006}. These
efforts demonstrated that exact analyses are indeed feasible even for
such large data sets as the WMAP data, which comprise several million
pixels. This is possible due to the very favorable scaling of the
Gibbs sampling algorithm.  While brute-force likelihood evaluations
scale as $\mathcal{O}(N_{\textrm{pix}}^3)$, $N_{\textrm{pix}}$ being
the number of pixels in the data set, the Gibbs sampler scales
identically to the map making operation.  For the special case of
uncorrelated noise and symmetric beams, this reduces further to
$\mathcal{O}(N_{\textrm{pix}}^{3/2})$.  Thus, even Planck-sized data
may be analyzed using these tools, as will be demonstrated in the
present paper.

Gibbs sampling thus provides an efficient route to the exact posterior
(or likelihood).  Moreover, it does not rely on any ad-hoc
approximations.  Even for the analysis of temperature data, this
proved to be both an important and subtle issue \citep{spergel:2006,
  eriksen:2006}. However, it is critical for polarization
measurements, because well-known approximate methods such as the
pseudo-$C_{\ell}$ methods (e.g., Chon et al.\ 2004) can lead to
aliasing of E-mode power into the much smaller B-mode power spectrum.
Although it is possible to construct ways around this problem
\citep{smith:2005}, exact methods such as full likelihood evaluations
or Gibbs sampling are clearly preferable solutions.  

We start by discussing the algorithms used for polarized Gibbs
sampling, extending the signal and power spectrum sampling steps from
temperature to polarization.  Then we analyze simulated data to verify
that the algorithm works and to determine the computational efficiency
of the method.

\clearpage

\section{Algorithms}
\label{sec:algorithms}

\subsection{Overview of Gibbs Sampling}

Gibbs sampling in the polarization case is essentially the same as in
the temperature case, with objects involved in the sampling re-defined
to account for the additional information.  For full details on the
methodology of Gibbs sampling as applied to CMB analysis, see
\citet{jewell:2004}; \citet{wandelt:2004}; and
\citet{eriksen:2004}.  

Specifically, the CMB signal is generalized to a vector of harmonics
coefficients $(a_{\ell m}^T, a_{\ell m}^E, a_{\ell m}^B)$ for each
$\ell$ and $m$, where the letters $T$, $E$, and $B$ stand for
temperature, electric/gradient, and magnetic/curl respectively.  The
covariance matrix $\BS$ of the CMB signal then becomes block-diagonal,
with an identical $3\times 3$ sub-matrix for each $m$ value at a
given $\ell$:
\begin{equation}
  \BC_\ell = \left(\begin{array}{ccc}
      C_\ell^{\textrm{TT}} & C_\ell^{\textrm{TE}} & C_\ell^{\textrm{TB}} \\
      C_\ell^{\textrm{TE}} & C_\ell^{\textrm{EE}} & C_\ell^{\textrm{EB}} \\
      C_\ell^{\textrm{TB}} & C_\ell^{\textrm{EB}} & C_\ell^{\textrm{BB}}
    \end{array}\right).
\end{equation}
The data are pixelized maps $\Bm$ of the Stoke's parameters
$I,Q,U$ of the form
\begin{equation}
  \Bm= \mathbf{A} \Bs+\Bn,
\end{equation}
where $\mathbf{A}$ is a linear operator that includes convolution with
an instrument beam and the transformation of the $T,E,B$ components of
the signal $\Bs$ into the Stokes parameters. Note that for the
rest of this paper, we will assume both the instrumental beam to be
symmetric and the noise $\Bn$ to be uncorrelated, having a diagonal
covariance matrix $\BN$. These are the reasons we can work with maps
instead of time-ordered data. However, to simplify the notation we
disregard in the following all issues concerning data format, beam
convolutions, multi-frequency observations etc., and model our data as
a simple sum of a signal term and a noise term. For the full
expressions, see Appendix~\ref{app:signal_sampling}.

Application of a galactic mask is implemented by increasing the noise
variance to infinity for masked pixels, or rather, by setting the
inverse noise covariance to zero. For full details, we refer the
interested reader to \citet{eriksen:2004}.

As in the temperature-only case discussed in \citet{jewell:2004} and
\cite{wandelt:2004}, we wish to sample from the $P(\BS | \Bd)$
posterior.  It is typically not easy to evaluate $P(\BS | \Bd)$
directly, because of a large and dense $(\BS + \BN)$ covariance
matrix, nor is it easy to sample from it directly. This is precisely
the motivation for Gibbs sampling, which allows sampling from a joint
density through the corresponding conditional densities. For the case
of CMB power spectrum estimation, this is done by first sampling from
$P(\BS, \Bs | \Bd)$ using $P(\BS | \Bs, \Bd)$ and $P(\Bs| \BS, \Bd)$,
(neither of which requires inversion of dense $(\BS + \BN)$ matrices),
and then marginalizing over $\Bs$. Using the fact that, given a
full-sky signal map the conditional density for the signal matrix is
independent of the data $P(\BS|\Bs,\Bd) = P(\BS|\Bs)$, the basic Gibbs
sampling scheme may be written in the following form,
\begin{eqnarray}
  \BS^{i+1} & \leftarrow & P(\BS|\Bs^i,\Bd) \label{sampleSpectrum}\\
  \Bs^{i+1} & \leftarrow & P(\Bs|\BS^{i+1}). \label{sampleSignal}
\end{eqnarray}
Here the symbol $\leftarrow$ indicates sampling from the distribution
on the right hand side. The only remaining problem is to establish the
correct sampling algorithms for each of the two conditional
distributions for polarized data, and this is the topic of the
following sections.

\begin{figure*}
\mbox{\epsfig{figure=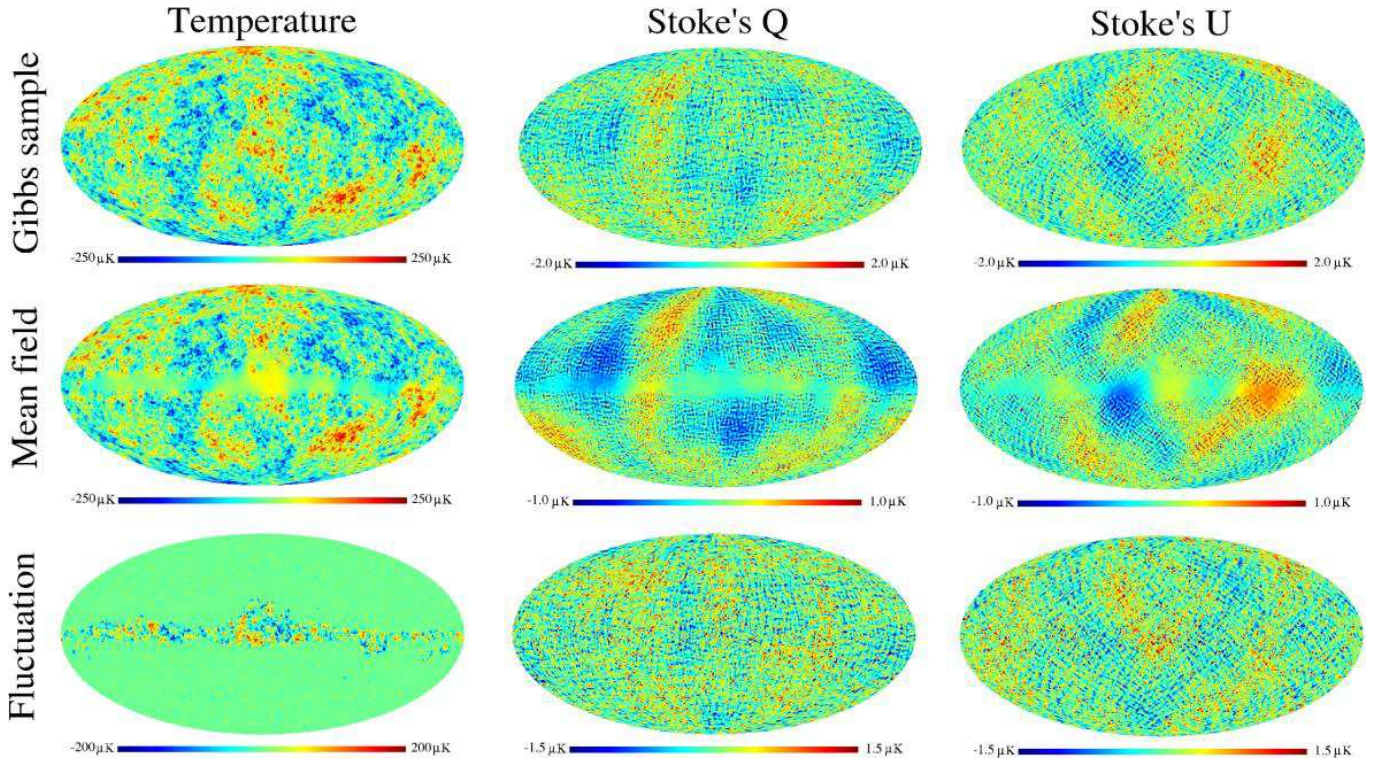,width=\linewidth,clip=}}  
\caption{Gibbs sampled signal maps. The three columns show, from left
    to right, temperature, Stoke's Q and Stoke's U parameters. The three
    rows show, from top to bottom, the complete Gibbs samples, the mean
    field (Wiener filtered) maps, and the fluctuation maps. The mean
    field map provides the information content of the data, and the
    fluctuation map provides a random complement such that the sum of
    the two is a full-sky, noiseless sky consistent both with the
    the current power spectrum and the data.}
  \label{fig:maps}
\end{figure*}

Note that if a continuous distribution for $P(\BS|\Bd)$ is desired, as
opposed to a set of individual samples, one may take advantage of the
known analytical form of the distribution $P(\BS|\Bs)$ by applying the
Blackwell-Rao estimator. This procedure was discussed in detail by
\cite{wandelt:2004} and \cite{chu:2005} for the temperature-only case,
and the generalization to polarization is once again straightforward.
The required modifications are written out in
Section~\ref{sec:blackwell_rao}.

\subsection{Signal Sampling}
\label{sec:signal_sampling}

The signal sampling equations for polarization are identical to those
for temperature-only data, taking into account the generalizations
mentioned above.  Specifically, the sky signal ($\Bs=\Bx+\By$) is
sampled (given the current covariance matrix $\BS$) by solving for the
mean field, $\Bx$, and fluctuation, $\By$, maps
\begin{eqnarray}
  \label{sampleMean}
  \left[\id+\BS^{1/2}\BN^{-1}\BS^{1/2}\right] \BS^{-1/2} \Bx &
  = & S^{1/2}\BN^{-1}\Bm,     \\
  \label{sampleFluctuation}
  \left[\id+ \BS^{1/2}\BN^{-1}\BS^{1/2}\right] \BS^{-1/2} \By  & 
  = & \mathbf{\xi} + \BS^{1/2} \BN^{-1/2}\mathbf{\chi},
\end{eqnarray}
where $\mathbf{\xi}$ and $\mathbf{\chi}$ are random maps containing
Gaussian unit variates (zero mean and unit variance) in each pixel for
each of the $I$, $Q$, and $U$ components\footnote{Note that
  $\BS^{1/2}$ and $\BS^{-1/2}$ must be symmetric for these equations
  to be valid. On the other hand, $\BN^{-1/2}$ only has to satisfy
  $\BN^{-1/2} (\BN^{-1/2})^T=\BN^{-1}$, and may be chosen to be the
  Cholesky decomposition.}. Note that the symbols in these equations
may be interpreted either in terms of pixel space or spherical
harmonic space objects. In practice, this is implemented in terms of
conversions between pixel and harmonic space with standard spherical
harmonics transforms. For example, the inverse noise covariance matrix
is given by $\mathbf{N}^{-1}$ in pixel space and $\mathbf{Y}^T
\mathbf{N}^{-1} \mathbf{Y}$ in harmonic space, where $\mathbf{Y}$ and
$\mathbf{Y}^T$ are the inverse and standard spherical harmonics
transforms, respectively. For explicit details on such computations,
see \citet{eriksen:2006}.

The signal sampling operation is by far the most demanding step of the
Gibbs sampler, because it requires the solution of a very large linear
system. Formally speaking, this corresponds to inverting a $\sim 10^6
\times 10^6$ matrix, which clearly is not computationally feasible
through brute-force methods. However, as described in detail by, e.g.,
\citet{eriksen:2004}, the systems in equations \ref{sampleMean} and
\ref{sampleFluctuation} may be solved by means of Conjugate Gradients
(CG). The computational scaling is thus reduced to the most expensive
step for applying the operator on the left hand side of the equations,
which for symmetric beams and uncorrelated noise is a standard
spherical harmonic transform.

The efficiency of the CG technique depends critically on the condition
number of the matrix under consideration. For our case, this is simply
the highest signal-to-noise ratio of any mode in the system.  As an
example, for a fixed pre-conditioner it takes about 60~iterations to
solve for the first-year WMAP data, about 120~iterations to solve for
the three-year WMAP data, and about 300~iterations to solve for the
Planck 100\,GHz data.

This is a particularly serious issue for CMB polarization
measurements. While these signatures by themselves have a very low
signal-to-noise ratio, and therefore should be easy to determine on
their own, the corresponding signal-to-noise ratio for temperature is
tremendous. Consequently, if a main goal is to estimate the TE
cross-spectrum, by far most of the CPU time is spent on temperature
map convergence. On the other hand, if all interest lies in E- and
B-modes, the temperature data may be disregarded completely (or
alternatively conditioned on by sampling from $P(a_{\ell m}^{E}, a_{\ell
m}^{B} | \Bd, a_{\ell m}^{T})$), and convergence is then achieved
rapidly even for CMBPol type missions. This will be explicitly
demonstrated in Section~\ref{sec:cmbpol}.

It is possible to reduce the computational expense of a CG search
significantly by pre-conditioning. One approach that has proved
successful so far is to pre-compute a subset of the coefficient matrix
in equations \ref{sampleMean} and \ref{sampleFluctuation}, and
multiply both sides of the equations by the inverted sub-matrix. Thus,
by inverting the most problematic parts of the matrix by
hand, the effective condition number is greatly reduced, and
significant speed-up may be achieved.

Currently, our pre-conditioner is constructed independently for the
temperature and polarization states. For the polarization components,
it is a diagonal matrix in EE and BB independently, while for the TT
correlations, it consists of a low-$\ell$ matrix that includes all
coefficients up to some $\ell_{\textrm{max}}$, and then the diagonal
elements at higher $\ell$'s \citep{eriksen:2004}. For WMAP-type
applications, we typically used $\ell_{\textrm{max}}=50$, which
requires 52\,MB of memory and about 1~minute of CPU time for
inversion. For upcoming Planck data, it will be desirable to
use a significantly larger pre-conditioner, and more realistic numbers
are $\ell_{\textrm{max}} \sim150$ or 200. This will require extensive
parallelization, and has not yet been implemented in our codes.  We
therefore still use a serial pre-conditioner up to
$\ell_{\textrm{max}} =70$ in this paper, and pay the extra cost in CG
iterations.

\begin{figure*}
\mbox{\epsfig{figure=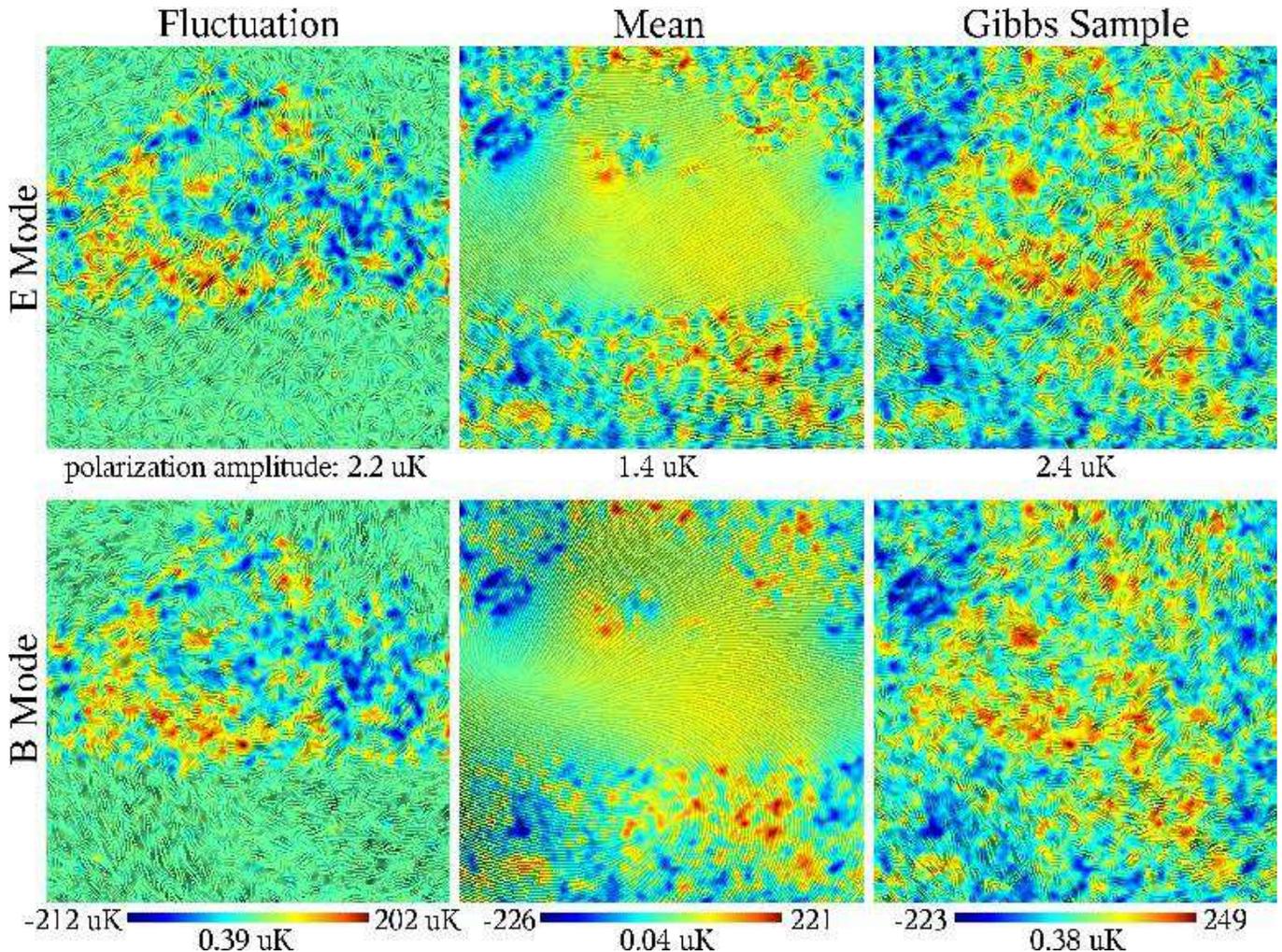,width=\linewidth,clip=}}  
  \caption{Close-up of the galactic center shown in Figure
    \ref{fig:maps}, emphasizing how the algorithm separates E- and 
    B-modes.  Each of the sampled maps (the sum of the fluctuation and
    mean field map) are full sky maps, so decomposing the polarization
    into E- and B-modes is straightforward.  The images show
    temperature as color and polarization overlayed as a fingerprint
    pattern of stripes.  The stripes are aligned with the direction of
    polarization.  They are darkest where the polarization is
    strongest, and they disappear where the polarization goes to zero
    \citep{cabral:1993}.  The maximum amplitude of the polarization is
    given in $\mu K$ and centered under each image.  The maps have
    been smoothed to 1 degree.  This is an orthogonal projection of
    the sky, about 60 degrees wide, centered on the Galactic center.
    The WMAP Kp0 galactic mask is visible in the fluctuation and mean
    terms.}
  \label{fig:alice}
\end{figure*}

\subsection{Power Spectrum Sampling}

Given the (full sky) signal polarization map sampled from $P(\Bs |
\BS, d)$ as described above, we must sample the signal covariance
matrix from $P(\BS | \Bs)$, which is explicitly given by
\begin{equation}
  P(\BS|\Bs) \propto \prod_\ell \frac{1}{\sqrt{\left|\BC_{\ell}
  \right|^{2\ell+1+2q}} }
  \exp\left(-\frac{1}{2}\,\mbox{tr}\,\sigma_\ell
  \BC_{\ell}^{-1}\right).
\label{eq:powspec_samp}
\end{equation}
Here we have assumed a prior of the form $P(\BS) \propto \prod_\ell
|\BC_\ell|^{-q}$ (i.e., $q=0$ for a uniform prior, and $q=1$ for a
Jeffreys prior), and we have defined
\begin{equation}
  \label{sigmaEllUnbinned}
  \sigma_\ell = \sum_{m=-\ell}^\ell \Bs_{\ell m} \Bs_{\ell m}^\dag.
\end{equation}
Each $\Bs_{\ell m}$ represents the signal in harmonic space, and is a
three dimensional complex-valued column vector containing the
coefficients for the T-, E-, and B-modes at that $\ell$ and $m$.  
This distribution is known as the inverse Wishart distribution
\citep{gupta:2000}.

Sampling from this conditional density can be done with a vector
generalization of the sampling algorithm described in
\citet{wandelt:2004}.  If $p\times p$ is the size of the matrix being
sampled (typically $p=3$ for polarization), then the required steps
for sampling are: 1) sample $n = 2\ell-p+2q$ vectors from a Gaussian
with covariance matrix $\sigma_{\ell}$; 2) compute the sum of outer
products of these independently sampled vectors; and 3) invert this
matrix. For full details on both the inverse Wishart distribution and
the sampling algorithm, we refer the interested reader to chapter 3 of
\citet{gupta:2000}.

There is a caveat for $\ell=2$. The Wishart distribution, from which
we derive our sampling algorithm, is defined only if $n \ge p$; if
not, the sampled matrix is singular. This is a problem for $\ell=2$
and a flat prior, since we would only sample one vector to form a
$3\times3$ matrix. Thus, the algorithm breaks down for this particular
case.  Fortunately, this is not a major problem in practice. Three
straightforward solutions are: 1) sample the $2\times 2$ TE block and
the B block of the matrix separately, assuming no TB or EB
correlations; 2) use a Jeffrey's prior ($q=1$); or 3) bin the
quadrupole and octopole together. Note that all other multipoles may
be sampled individually by the above algorithm without modifications.

\paragraph{Binning}
\label{sec:binning}

As discussed by \citet{eriksen:2006}, it is highly desirable for the
Gibbs sampler to be able to bin several power spectrum multipoles
together. The main advantage of this is improved sampling efficiency:
As currently implemented, the step size taken between two consecutive
Gibbs samples is given by cosmic variance alone. The full posterior,
however, is given by both cosmic variance and noise. Therefore, in the
low signal-to-noise regime, one must take a larger number of steps to
obtain two independent samples. The easiest way of improving on this
is simply to bin many multipoles together, and thereby increase the
signal-to-noise ratio of the power spectrum coefficient. In practice,
we choose bins such that the signal-to-noise ratio is always larger
than some limit, say 3.

Since the CMB power spectrum is roughly proportional to
$1/\ell(\ell+1)$, it is convenient to define uniform bins in $C_\ell
\ell(\ell+1)$.  We therefore redefine $\sigma_\ell$ for bin $b =
[\ell_{\textrm{min}},\ell_{\textrm{max}}]$ as
\begin{equation}
  \label{binSigma}
  \sigma_\ell = \sum_{\ell \in b} \sum_{m=-\ell}^\ell 
  \ell(\ell+1) \Bs_{\ell m} \Bs_{\ell m}^\dag.
\end{equation}

Note that there are now
\begin{equation}
  M = \sum_{\ell \in b} (2\ell+1) = (\ell_{\textrm{max}}+1)^2 -
  \ell_{\textrm{min}}^2
\end{equation}
independent spherical harmonic modes contributing to this power
spectrum coefficient.  Thus, the inverse Wishart distribution has $n =
M-p-1+2q$ degrees of freedom rather than $n = 2\ell-p+2q$. With this
modification, the basic sampling algorithm remains unchanged, but
since we have sampled $\BC_{b} = \ell(\ell+1) \BC_\ell$ and not
$\BC_{\ell}$, the actual power spectrum coefficients are given by
$\BC_\ell = \BC_{b} / \ell(\ell+1)$ for each $\ell$ in bin $b$.

\subsection{Separation of E- and B-modes}

We now make a brief comment on the so-called E-B coupling problem that
plagues most approximate methods, such as the pseudo-$C_{\ell}$
methods (see, e.g., Smith 2005). Briefly put, the problem lies in the
fact that the spherical harmonics are not orthogonal on a cut sky, and
this may result in leakage from the (much larger) E-mode power into
the B-mode power spectrum.

Exact methods such as exact likelihood analyses or Gibbs sampling do
not have this problem. This may be understood intuitively in terms of
the signal sampling process illustrated in Figures \ref{fig:maps} and
\ref{fig:alice}.  Obtaining a complete sky sample for the Gibbs
sampler is a two step process. First, one filters out as much
information as possible from the observed data using a Wiener
filter. Second, one replaces the lost power due to noise and partial
sky coverage by a random fluctuation term. The sum of the two is a
{\it full-sky\/}, noiseless sample that is consistent with the
data. Because it is a full-sky sample, no E-B coupling arises.

\subsection{Blackwell-Rao Estimator}
\label{sec:blackwell_rao}

The Gibbs sampler provides a set of samples of the signal covariance
matrix $\BS$.  In practice, it is often preferable to have a smooth
description of the probability density of $\BS$.  In such cases, one
can use the Blackwell-Rao estimator, which takes advantage of the
known analytical form of the probability distribution $P(\BS|\Bs)$ and
uses the set of signal samples $\{\Bs\} = \{\Bs^1,\ldots,\Bs^k\}$ to
approximate $P(\BS|\Bd)$.

An intuitive understanding of the Blackwell-Rao estimator may be found
in terms of the usual Gibbs sampling algorithm. Within the theory of
Gibbs sampling (or more generally Markov Chain Monte Carlo), it is
perfectly valid to sample one parameter more often than others, so
long as the sampling scheme is independent of the current ``state'' of
the Markov chain.  In particular, one may choose to sample $\BS$ one
thousand times for each time one samples $\Bs$, and thereby obtain
more power spectrum samples (although not sky signal samples) with
negligible cost. The result is a smooth power spectrum histogram. The
Blackwell-Rao estimator takes this idea to the extreme, and replaces
the power spectrum sampling step by the corresponding analytical
distribution. The result is a highly accurate and smooth description
of $P(\BS|\Bd)$ that is very useful for, say, estimation of
cosmological parameters \citep{wandelt:2004, chu:2005, eriksen:2006}.

For full details on this estimator for the temperature-only case, we
refer the interested reader to \citet{chu:2005}. But again, the
generalization to polarization is indeed straightforward, and the
generalized estimator reads
\begin{equation}
  P(\BS|\{\Bs\}) \propto 
\sum_j \prod_\ell \frac{\sqrt{|\sigma_\ell|}^{2\ell+2q-p}}
{\sqrt{|\mathbf{C}_\ell|}^{2\ell + 1 + 2q}} 
\exp\left(-\frac12 \text{tr}\, \sigma_\ell^j \mathbf{C}_\ell^{-1}
\right)
\end{equation}
where $\BC_\ell$ is the $3\times3$ sub-matrix of $\BS$ for a given
  $\ell$, and $j$ is the index over Gibbs samples.

\section{Application to simulated data}

We now apply the methodology described in Section~\ref{sec:algorithms}
to simulated data. Two different cases are considered to highlight
different features. In the first, we consider a low-resolution, high
signal-to-noise ratio experiment aimed at detecting primordial
B~modes. The main goal of this exercise is to demonstrate the fact
that the so-called E/B coupling problem that plagues approximate
methods is not an issue for exact methods. Second, we consider a
high-resolution simulation based on the Planck 100\,GHz channel to
demonstrate that Gibbs sampling is feasible even for very large CMB
data sets.

\subsection{Low-resolution B-mode experiment (CMBPol)}
\label{sec:cmbpol}

\begin{figure}
\mbox{\epsfig{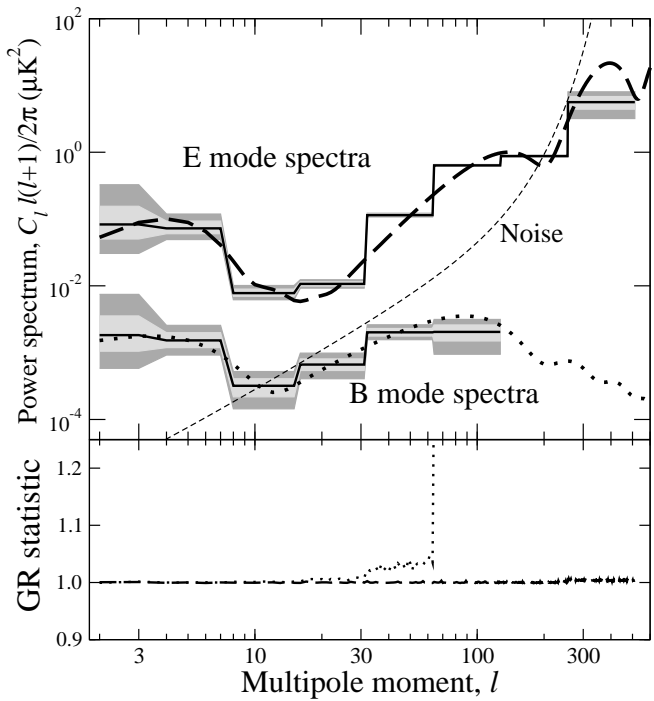}}
\caption{Reconstructed E- and B-mode power spectra from the
    low-resolution analysis. Input spectra are shown as dashed and
    dotted lines, respectively, while the reconstructed posterior
    distributions are indicated by solid curves (posterior maximum) and
    gray regions (one and two sigma confidence regions). The
    corresponding noise spectrum is given by a thin dashed line. The
    Gelman-Rubin convergence statistic as a function of multipole is
    shown in the bottom panel. }
  \label{fig:lowl_spectra}
\end{figure}

\begin{figure}
\mbox{\epsfig{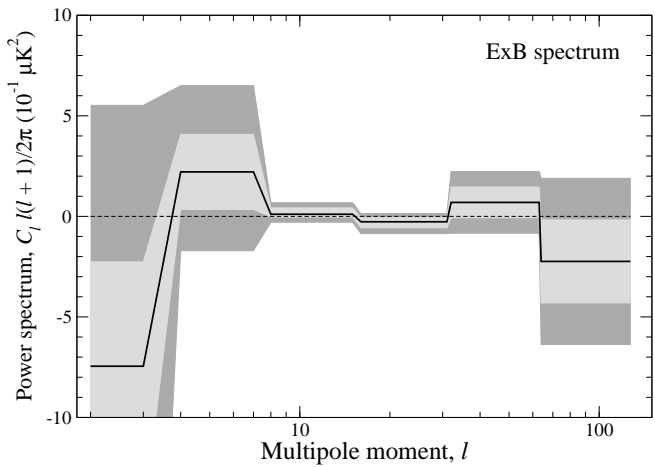}}
\caption{The E$\times$B cross-spectrum from the low-resolution
    analysis.}
  \label{fig:ExB_spectrum}
\end{figure}

Our first case corresponds to a possible future mission targeting the
primordial B-modes that arise during the inflationary period. Such
modes are expected to have a very low amplitude and to be limited to
large angular scales.  Some case studies for a B-mode mission
therefore emphasize extreme sensitivity over angular resolution, and
we adopt similar characteristics for this exercise.

As discussed in Section~\ref{sec:signal_sampling}, the convergence
ratio for the conjugate gradient search depends critically on the
signal-to-noise ratio of the data. In order to achieve acceptable
performance when analyzing temperature observations with the
sensitivity required for detecting B-modes, a much better
pre-conditioner than what we have currently implemented is
required. We therefore only consider the E- and B-mode spectra here,
and not the temperature spectrum.

The simulated data set consists of the sum of a CMB component and a
white noise component. The CMB realization was drawn in harmonic space
from a Gaussian distribution with a $\Lambda$CDM spectrum (downloaded
from the WMAP3 parameter table at LAMBDA) having a tensor contribution
of $r\simeq 0.03$. Multipoles up to $\ell_{\textrm{max}} = 512$ were
included. This realization was then convolved with a $1^{\circ}$ FWHM
Gaussian beam and $N_{\textrm{side}}=256$ pixel window, and projected
onto a HEALPix\footnote{http://healpix.jpl.nasa.gov/} grid. Next,
uniform (and uncorrelated between $Q$ and $U$) noise of
$1\,\mu\textrm{K}$ rms was added to each pixel. Finally, the WMAP3
polarization mask
\citep{page:2006} was applied, removing 26.5\% of the sky from the
analysis.

We adopted a binning scheme logarithmic in $\ell$, such that $b_i =
[2^i, 2^{i+1}-1]$. Note that this is not directly connected to the
signal-to-noise ratio of the data themselves, and this will have
consequences for the convergence properties of the high-$\ell$ B-mode
bins.  However, our main focus in this paper is the method itself, and
this scheme is chosen to illustrate the effect of both high and low
signal-to-noise binning, not to obtain an optimal power spectrum.

The simulation was then analyzed with the Gibbs sampler described
earlier, producing 1000~sky samples in each of five independent Markov
chains. The CPU cost for producing one sample was 10~minutes, or a
wall clock time of 2.5~minutes when parallelized over four
processors. The total running time was thus 42~hours using
20~processors.  For each sky sample, 20~independent power spectrum
samples were drawn in order to obtain smoother $C_{\ell}$ confidence
regions. (See the discussion of the Blackwell-Rao estimator in
Section~\ref{sec:blackwell_rao} for more details.)

We first consider the reconstructed auto-spectra, which are shown in
the top panel of Figure~\ref{fig:lowl_spectra}. The input (unbinned)
spectra are given by dashed and dotted lines for E- and B-modes,
respectively, and the reconstructed (binned) posterior maximum spectra
are shown by solid black lines. One and two sigma confidence regions
are marked by gray regions. Finally, the beam deconvolved noise
spectrum is indicated by a thin dashed line.

In the bottom panel, we show the Gelman-Rubin convergence statistic
\citep{gelman:1992} as computed from the $\sigma_{\ell}$ sky signal
power spectra for each $\ell$. This is much more conservative than
computing the same statistic from the $C_{\ell}$ samples, for two
reasons.  First, convergence in the binned power spectrum is achieved
faster than convergence in each sky mode. Second, cosmic variance only
contributes to the power spectrum and not the signal on the sky.
Therefore, this may be accounted for either analytically through the
Blackwell-Rao estimator or by re-sampling the $C_{\ell}$ spectra given
the $\sigma_{\ell}$'s. In other words, a small error in the sky signal
variance does not affect the full posterior significantly if the
desired distribution is anyway dominated by cosmic variance.

A general recommendation is that the Gelman-Rubin statistic, $R$,
should be less than 1.1 or 1.2 to claim convergence, although the
value depends on the particular application and initialization
procedure, and should be compared against other methods such as
jack-knife tests. However, for the particular case shown in
Figure~\ref{fig:lowl_spectra}, it is clear that the E-mode spectrum
has converged very well everywhere, while the B-mode spectrum only has
converged up to $\ell \approx 60$.

\begin{figure*}
\mbox{\epsfig{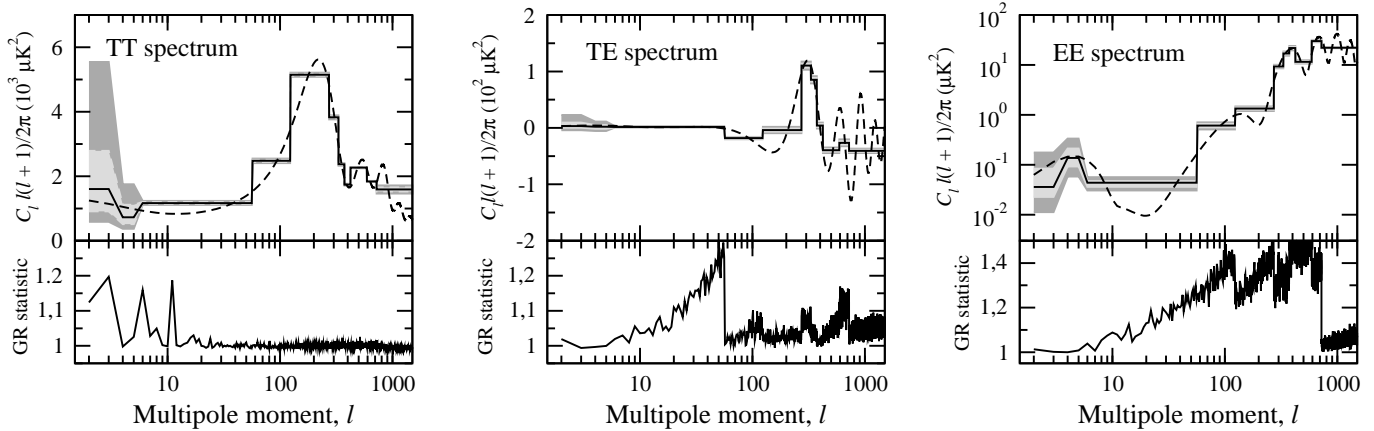}}
\caption{Reconstructed power spectra from the high-resolution Planck
    100 GHz simulation. The true spectra are shown as dashed lines,
    and the reconstructed posterior distributions are given by a
    maximum posterior value (solid lines) and 68 and 95\% confidence
    region. The Gelman-Rubin convergence statistics are shown in the
    bottom panels.}
  \label{fig:highres_spectra}
\end{figure*}

As discussed in Section~\ref{sec:binning}, this behavior can be
understood intuitively in terms of signal-to-noise ratio. Since the
step size between two signal samples is given by cosmic variance
alone, while the full posterior distribution is given by both cosmic
variance and noise, it takes a large number of Gibbs steps to diffuse
efficiently in the very low signal-to-noise regime. Further, the noise
spectrum is about three orders of magnitude larger than the B-mode
spectrum at $\ell \gtrsim 100$, and the Gibbs sampler is therefore
unable to probe the full distribution with a reasonable number of
samples.

To resolve this issue, we bin the power spectrum. However, the binning
scheme was not tuned to obtain constant signal-to-noise in each bin,
but was rather arbitrary. The result is clearly seen in the
Gelman-Rubin statistic: For $\ell \lesssim 60$ the signal-to-noise per
bin is high, and convergence is excellent. At $\ell \gtrsim 60$, it is
low, and the convergence is very poor. The way to resolve this would
have been to choose larger bins at higher $\ell$'s.

In Figure~\ref{fig:ExB_spectrum}, we show the E$\times$B
cross-spectrum. As expected, this is nicely centered on zero.

We end this section by commenting on the applicability of this
formalism to a possible future CMBPol type mission. As is well known,
the main problems for such a mission will not be primarily statistical
issues of the type discussed above, but rather systematics in various
forms. Two important examples are correlated noise and asymmetric
beams. However, if it is possible to pre-compute the complete
$\mathbf{A}^{\textrm{T}} \mathbf{N}^{-1} \mathbf{A}$ matrix for the
data set under consideration, then these two important effects may be
fully accounted for using the methods described here.  And for a
low-resolution CMBPol mission this may be possible.  For an upper
multipole limit of, say, $\ell_{\textrm{max}} = 300$ there is a total
of 2$\times$90\,000 = 180\,000 polarized spherical modes to account
for. In other words, one has to store and invert a $180\,000\times
180\,000$ matrix in order to analyze such an experiment
exactly. Although this is a considerable computational problem, it is
quite tractable already with current computers. Thus, if it is
possible to compute this matrix in the first place for a given
experiment, an exact and complete analysis is feasible using the
methods described in this paper.

\subsection{High-resolution T+E experiment (Planck)}

In order to demonstrate the feasibility of this method for analyzing
even the largest planned data set, we now consider a simulation with
properties similar to those of the Planck 100\,GHz
instrument. Specifically, the grid resolution is chosen to be
$N_{\textrm{side}} = 1024$ (corresponding to a $3.4\arcmin$ pixel
size), the maximum multipole moment is $\ell_{\textrm{side}} = 1500$,
the beam size is $9.5\arcmin$, and the noise level is
$38.2\,\mu\textrm{K}$ RMS per pixel for temperature and
$61\,\mu\textrm{K}$ per Q/U pixel. These noise levels are a factor of
two higher than the goal levels for the 100\,GHz channel, given in the
Planck
bluebook\footnote{http:$//$www.rssd.esa.int$/$SA$/$PLANCK$/$docs$/$Bluebook-ESA-SCI(2005)1$\_$V2.pdf}. No
noise correlations between $T$, $Q$ and $U$ were included, the B-mode
spectrum was set to zero, and the sky cut was chosen to be the WMAP
Kp2 mask.

Such large data sets are certainly a challenge for the Gibbs sampling
algorithm, and the computational requirements are
considerable. Specifically, the CPU time for generating one sample
(requiring $\sim250$--300 CG iterations) is about 16\,CPU hours when
using the low-$\ell$ preconditioner described by \citet{eriksen:2004}
up to $\ell=70$. Better preconditioners will of course reduce this
cost significantly.

However, it is important to note that even though this is an expensive
operation, it is by no means prohibitive. To obtain a reasonably well
converged posterior distribution, one requires on the order of
$\sim10^3$ independent samples, and this would then require $\sim10^4$
CPU hours. Of course, this number must be multiplied with a
significant factor for an actual production analysis (e.g, number of
frequency bands or data combinations), but considering the tremendous
efforts spent on obtaining the Planck data in the first place, this
amount of CPU time is a most reasonable cost for analyzing them.

For the high-resolution analysis presented in this paper, we produced
a total of 800~sky samples, divided over eight independent
chains. Again, 20~independent power spectrum samples were then drawn
from each of these for visualization purposes. The results from these
computations are summarized in Figure~\ref{fig:highres_spectra},
showing both the reconstructed power spectra and the corresponding
convergence statistics.

With the chosen binning scheme and number of samples, we see that the
TT spectrum has converged well everywhere, while the TE spectrum has
some small problems at the end of the second bin. The EE spectrum
would clearly have benefited from more samples, and even more
importantly, slightly larger bins; increasing the bin size by, say,
20\% would have resolved both the TE and EE issues.

However, as far as computational feasibility goes, the important part
is the signal sampling step, and not binning or re-sampling issues;
these can always be adjusted given some crude knowledge of the data
set under consideration. Therefore, the fact that already this first
implementation of the polarized Gibbs sampler is able to produce
hundreds of sky samples with only a few days on a standard computer
cluster, is a direct demonstration of computational feasibility for
even Planck-sized data sets.

\section{Conclusions}

This paper extends the Gibbs sampling technique to polarized power
spectrum estimation.  We have detailed the necessary
generalization steps relative to the original temperature-only
descriptions given by \citet{jewell:2004}, \citet{wandelt:2004} and
\citet{eriksen:2004}, and have considered computational aspects of
polarized analysis.

The algorithm was demonstrated with two specific examples. First,
considering a possible CMBPol type mission, we showed that the Gibbs
sampler cleanly separates E- and B-modes, and no special care is
required. This is in sharp contrast to approximate methods such as
so-called the pseudo-$C_{\ell}$, for which great care must be taken in
order for the larger E-modes not to compromise the minute B-modes.

Second, we analyzed a Planck-sized data set, demonstrating that the
algorithm is useful for analyzing the quantity of data which will come
from near-future CMB experiments.

The Gibbs sampling results presented here use symmetric beams and
noise which is uncorrelated between pixels. However, the Gibbs
sampling algorithm has potential to analyze considerably more
complicated data sets than these. For Planck, the solution lies in
exploiting the very regular scanning strategy, which reduces the
computational burden of a time-ordered data analysis. For a future
CMBPol mission, the solution lies in the relatively large angular
scales required. Since it is possible to invert the noise covariance
matrix for multipoles up to several hundreds, one may pre-compute the
all-important $\mathbf{A}^T \mathbf{N}^{-1}\mathbf{A}$ matrix. After
paying this high one-time cost, efficient and exact analysis is
feasible using the methods described in this paper.

Finally, we re-emphasize that the Gibbs sampler provides a direct
route to the exact likelihood (and to the Bayesian posterior), and it
is much more reliable than approximate methods. This issue has been
demonstrated explicitly through the analysis of the three-year WMAP
data, where an approximate likelihood between $\ell=13$ and 30 caused
a non-negligible bias in the spectral index $n_{\textrm{s}}$
\citep{eriksen:2006}. Using Gibbs sampling, such worries are greatly
reduced. Further, this paper demonstrates that the method is in fact
capable of analyzing the amount of data that will come from the Planck
mission with reasonable computational resources. It therefore seems
very likely that this method will play a significant role in the
analysis of future Planck data.

\begin{acknowledgements}
  We thank Kendrick Smith, Graca Rocha and Charles Lawrence for useful
  and interesting discussions. We acknowledge use of the HEALPix
  software \citep{gorski:2005} and analysis package for deriving the
  results in this paper. We acknowledge use of the Legacy Archive for
  Microwave Background Data Analysis (LAMBDA). This work was partially
  performed at the Jet Propulsion Laboratory, California Institute of
  Technology, under a contract with the National Aeronautics and Space
  Administration. HKE acknowledges financial support from the Research
  Council of Norway.  BDW and DLL acknowledge support through NSF
  grant AST-0507676 and NASA JPL subcontract 1236748.
\end{acknowledgements}

\appendix

\section{Signal Sampling}
\label{app:signal_sampling}

\newcommand{\proj}{\mathbf{P}}

Equations \ref{sampleMean} and \ref{sampleFluctuation} have been
written as simply as possible, for clarity when describing the
extension of our Gibbs sampling algorithm to polarization.  A more
realistic treatment will involve multiple channels, symmetric beams,
the pixel window function, and a cutoff at some value of $\ell$.
In this appendix we write out, for reference, the log likelihood for
$\Bs$ and the sampling equations that one derives from this.

Let the index $i$ run over channels.  Let $\BB_i$ be the beam
smoothing function, and $\BW$ be the HEALPix pixel window smoothing
function.  If all channels are at the same resolution, then there is
only one pixel window function; otherwise $\BW$ will need an $i$ index
as well.  Let $\proj$ be a projection operator that removes all modes
with $\ell$ above some cutoff.  Note that $\proj$, $\BW$, and
$\BB_i$ all commute, and $\proj$ commutes with $\BS$.  
As before, $\Bm_i$ are the maps and $\Bs$ is the
signal.  For generality, we also include a foreground component
$\Bf_i$, which is not otherwise discussed in this paper.
{\small
\begin{equation}
  -2 \log P(\Bs|\BS, \Bm_i, \Bf_i, \BN_i, \BB_i, \BW) =  \Bs^T \BS^{-1} \Bs + 
  \sum_i \left(\Bm_i - \BB_i \BW \Bs - \BB_i \BW \Bf_i\right)^T \proj \BN_i^{-1} \proj 
  (\Bm_i - \BB_i \BW \Bs - \BB_i \BW \Bf_i) + \mbox{const.}
\end{equation}
}
From the above equation, it is clear that $\BW$ can be absorbed into $\BB_i$, so we do
this and drop $\BW$ from the equations.  The equations for sampling $\Bs = \Bx + \By$
become:
\begin{eqnarray}
  \left[\proj+\proj \BS^{1/2}\sum_i(\BB_i\proj\BN^{-1}\proj\BB_i)\BS^{1/2}\proj\right] \BS^{-1/2} \proj \Bx &
  = & \proj \BS^{1/2}\sum_i \BB_i \proj \BN_i^{-1}\proj (\Bm_i - \BB_i \Bf_i)\\
  \left[\proj+\proj \BS^{1/2}\sum_i(\BB_i\proj\BN^{-1}\proj\BB_i)\BS^{1/2}\proj\right] \BS^{-1/2} \proj \By &
  = & \proj \mathbf{\xi} + \proj \BS^{1/2} \sum_i \BB_i \proj \BN_i^{-1/2}\mathbf{\chi}_i,
\end{eqnarray}
where now we have several maps $\mathbf{\chi}_i$ of Gaussian unit
variates.  Recall that these equations require the square root of
$\BS$ to be symmetric.


\begin{thebibliography}{}

\bibitem[Barkats et al.(2005)]{barkats:2005} Barkats, D., et al.\ 
2005, \apjl, 619, L127

\bibitem[Cabral and Leedom(1993)]{cabral:1993} Cabral, B. \& Leedom,
  L.\ 1993, Proceedings of the 20th annual conference on Computer
  graphics and interactive techyniques, p.\ 263--270, ACM Press

\bibitem[Chon et al.(2004)]{chon:2004} Chon, G., Challinor, A.,
Prunet, S., Hivon, E., \& Szapudi, I.\ 2004, \mnras, 350, 914

\bibitem[Chu et al.(2005)]{chu:2005} Chu, M., Eriksen, H.~K., Knox,
  L., G{\'o}rski, K.~M., Jewell, J.~B., Larson, D.~L., O'Dwyer, I.~J.,
  \& Wandelt, B.~D.\ 2005, \prd, 71, 103002

\bibitem[Eriksen et al.(2004)]{eriksen:2004} 
  Eriksen, H.~K., et al.\ 2004, \apjs, 155, 227

\bibitem[Eriksen et al.(2006)]{eriksen:2006} 
  Eriksen, H.~K., et al.\ 2006, \apj, 641, 665

\bibitem[Gelman \& Rubin(1992)]{gelman:1992}
  Gelman, A., \& Rubin, D. 1992, Stat. Sci., 7, 457

\bibitem[G{\'o}rski et al.(2005)]{gorski:2005} 
  G{\' o}rski, K.~M., Hivon, E., Banday, A.~J., Wandelt, B.~D.,
  Hansen, F.\,K., Reinecke, M., \& Bartelmann, M. 2005, \apj, 622, 759

\bibitem[Gupta \& Nagar(2000)]{gupta:2000}
  Gupta, A.~K. \& Nagar, D.~K. 2000, Matrix Variate Distributions

\bibitem[Jewell et al.(2004)]{jewell:2004} 
  Jewell, J., Levin, S., \& Anderson, C.  H.  2004, \apj, 609, 1

\bibitem[Kuo et al.(2004)]{kuo:2004} Kuo, C.~L., et al.\ 2004, \apj,
  600, 32

\bibitem[Leitch et al.(2002)]{leitch:2002} Leitch, E.~M., et al.\
2002, \nat, 420, 763

\bibitem[Leitch et al.(2005)]{leitch:2005} Leitch, E.~M., Kovac,
J.~M., Halverson, N.~W., Carlstrom, J.~E., Pryke, C., \& Smith,
M.~W.~E.\ 2005, \apj, 624, 10

\bibitem[Montroy et al.(2005)]{montroy:2005} Montroy, T.~E., et al.\ 
  2005, ApJ, submitted, [astro-ph/0507514]

\bibitem[O'Dwyer et al.(2004)]{odwyer:2004} 
  O'Dwyer, I. J. et al.\ \apjl, 617, L99

\bibitem[Page et al.(2006)]{page:2006} Page, L., et al.\ 2006, 
  ApJ, submitted, [astro-ph/0603450]

\bibitem[Piacentini et al.(2005)]{piacentini:2005} 
  Piacentini, F, et al.\ 2005, ApJ, submitted [astro-ph/0507507]



\bibitem[Readhead et al.(2004)]{readhead:2004} Readhead, A.~C.~S., et
al.\ 2004, Science, 306, 836

\bibitem[Smith(2005)]{smith:2005} Smith, K.~M.\ 2005, Phys.\ Rev.\ D,
  submitted, [astro-ph/0511629]
  
\bibitem[Spergel et al.(2006)]{spergel:2006} Spergel, D.~N., et al.\ 
  2006, ApJ, submitted, [astro-ph/0603449]
  
\bibitem[Wandelt et al.(2004)]{wandelt:2004} 
  Wandelt, B.~D., Larson, D.~L., \& Lakshminarayanan, A.\ 2004, \prd,
  70, 083511


\end{thebibliography}
\end{document}